\documentclass[prl,amssymb,twocolumn,tightenlines,showpacs]{revtex4-1}

\usepackage{graphicx}
\usepackage{amsmath}
\usepackage{verbatim}
\usepackage{subfigure}

\usepackage{float}

\usepackage{epstopdf}
\usepackage{color}
\usepackage{natbib}

\definecolor{mygrey}{gray}{0.35}
\definecolor{myblue}{rgb}{0.2,0.2,0.8}
\definecolor{myzard}{cmyk}{0,0,0.05,0}
\definecolor{mywhite}{rgb}{1,1,1}
\definecolor{myred}{rgb}{1,0.,0.3}

\usepackage[colorlinks=true,citecolor=myblue,linkcolor=myred,urlcolor=myblue]{hyperref}

\begin{document}

\title{Simple source for large linear cluster photonic states}

\author{Y. Pilnyak}
\affiliation{Racah Institute of Physics, Hebrew University of
Jerusalem, Jerusalem 91904, Israel}
\author{N. Aharon}
\affiliation{Racah Institute of Physics, Hebrew University of
Jerusalem, Jerusalem 91904, Israel}
\author{D. Istrati}
\affiliation{Racah Institute of Physics, Hebrew University of
Jerusalem, Jerusalem 91904, Israel}
\author{E. Megidish}
\affiliation{Racah Institute of Physics, Hebrew University of
Jerusalem, Jerusalem 91904, Israel}
\author{A. Retzker}
\affiliation{Racah Institute of Physics, Hebrew University of
Jerusalem, Jerusalem 91904, Israel}
\author{H. S. Eisenberg}
\affiliation{Racah Institute of Physics, Hebrew University of
Jerusalem, Jerusalem 91904, Israel}


\begin{abstract}
The experimental realization of many-body entangled states is one
of the main goals of quantum technology as these states are a key
resource for quantum computation and quantum sensing. However,
increasing the number of photons in an entangled state has been
proved to be a painstakingly hard task. This is a result of the
non-deterministic emission of current photon sources and the
distinguishability between photons from different sources.
Moreover, the generation rate and the complexity of the optical
setups hinder scalability. Here we present a new scheme that is
compact, requires a very modest amount of components, and avoids
the distinguishability issues by using only one single-photon
source. States of any number of photons are generated with the
same configuration, with no need for increasing the optical setup.
The basic operation of this scheme is experimentally demonstrated
and its sensitivity to imperfections is considered.
\end{abstract}

\maketitle


The majority of quantum information protocols require entanglement between
different subsystems of a quantum state\cite{Zhao04}. For
example, the demonstration of a useful quantum computer, will
involve the controlled entanglement of probably thousands of
quantum elements\cite{Nielsen04}. This has been
proved so far to be a very hard task, especially with photons
whose generation rates decrease exponentially with their
numbers\cite{Yao12}. This task is simplified when several single photon
sources are combined on an optical chip through integrated
waveguides\cite{Matthews09}, but the required complete
indistinguishability between all of the sources is hard to
achieve\cite{Maunz07,Lettow10,Flagg10,Patel10,Bernien12,Portalupi15}.
The highest numbers of entangled photons were created by combining
polarization entangled photon pairs from several parametric
down-conversion (PDC) sources\cite{Yao12}. Nevertheless,
in order to increase the number of entangled photons, more
sources, entangling operations, and matching delay lines are
required. In addition, extremely low state detection rates were
observed, due to the probabilistic nature of PDC, where typical
photon-pair generation probabilities are of few percents.
Previously, we demonstrated an approach that simplified this
setup, such that only a single entangled photon pair source and a
single delay line are required, regardless of the size of the
generated state\cite{Megidish12}. This setup still suffers from
low generation rates, as it also uses PDC.

In this work we present a new scheme that generates multi-photon
linear cluster states\cite{Walther05}.
The scheme is using single photon sources\cite{Orrit05}, where
on-demand operation is almost
achieved\cite{Claudon10,Liebermeister14,Sapienza15,Somaschi15},
and a single delay line in a loop arrangement. Thus,the amount of
resources is reduced to only one single-photon emitter and one
entangling gate. Such delay loops were previously used in schemes
where quantum information is time-bin
encoded\cite{Schreiber12,Rohde15}. The use of only a single source
simplifies the efforts for indistinguishability, removing the need
for fine and stable spectral tuning of the different
sources\cite{Santori02,Gazzano13,Wei14,Portalupi15}.
Nevertheless, consequent emissions still require
indistinguishability. The typically long coherence times of such
sources compared to PDC sources, imply even lower sensitivity to
the length of the delay line. Furthermore, these sources' narrow
spectral widths enable the use of optical fibers, which can not be
used in the case of PDC photons due to dispersion induced
distinguishability. This scheme requires a photon-photon
entangling operation. Although there are many efforts to
demonstrate a device with such interaction, there is still no
practically satisfying
demonstration\cite{Volz12,Englund12,Chen13,Baur14,Shomroni14}.
Currently, the most efficient way to entangle the polarization of
two photons is by post-selection of the state after passing
through a polarizing beam-splitter (PBS)\cite{Pan01}. The
post-selection step introduces a 50\% success probability, but
relatively high fidelities. Yet, whichever is the entangling
process that is used, our scheme only requires such a single
device.

\begin{figure}[t]
\includegraphics[angle=0,width=86mm]{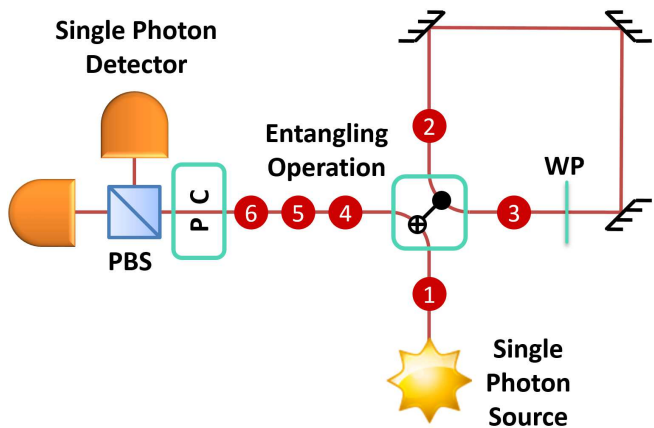}
\caption{\label{Fig1}\textbf{A scheme for generating multi-photon
cluster states.} A single-photon source emits photons
consecutively into an optical path, leading into a Sagnac-like
loop. Inside the loop there is a half-wave plate (WP) at
$22.5^\circ$, and its delay time is matched to the time between
consecutive photon emissions. At the point of intersection of the
optical paths, a photon entangling operation entangles one photon
from the source and another from the delay line. One photon
continues into the delay loop and the other one further towards a
polarization controller (PC), a polarizing beam splitter (PBS),
and detectors. The state is characterized by different detection
sequences. See main text for more details.}
\end{figure}

The main principles of our scheme are presented in Fig.
\ref{Fig1}. A single photon source emits a polarized photon into
position 1. The photon reaches a CNOT entangling gate, but passes
it through with no effect into the delay loop at position 2.
Inside the loop, the photon is rotated by $45^\circ$ with a
half-wave plate and arrives to position 3. Synchronized with the
second arrival of this photon to the entangling gate, the single
photon source is triggered again such that the two photons are
simultaneously entering both input ports of the entangling gate at
positions 1 (target) and 3 (control). The CNOT gate generates
entanglement between the photon that has left the loop and the
other one that is still inside it. After the photon inside the
loop is rotated by the wave plate, the source is synchronously
triggered for the third time. This time, the entangling operation
combines the new photon with the previous state and creates a
three photon GHZ state\cite{Greenberger90}
Continuing this process successively, results in an ever growing
chain of entangled photons in a linear cluster state. The delay
loop serves as a quantum memory that connects between generations,
in a similar way to the entangled single-photon quantum dot source
of the proposal in Ref. 30.

The output stream of photons is analyzed at different polarization
bases by rotating the photon polarizations with a polarization
controller and detecting the photons at the output ports of a PBS.
By using fast Pockels cells, it is possible to change the
projection direction of some photons according to the previous
measurement outcomes of others. This procedure is called 'feed
forward', and it is required for quantum computation protocols
using graph states\cite{Prevedel07}.

\begin{figure}[t]
\includegraphics[angle=0,width=86mm]{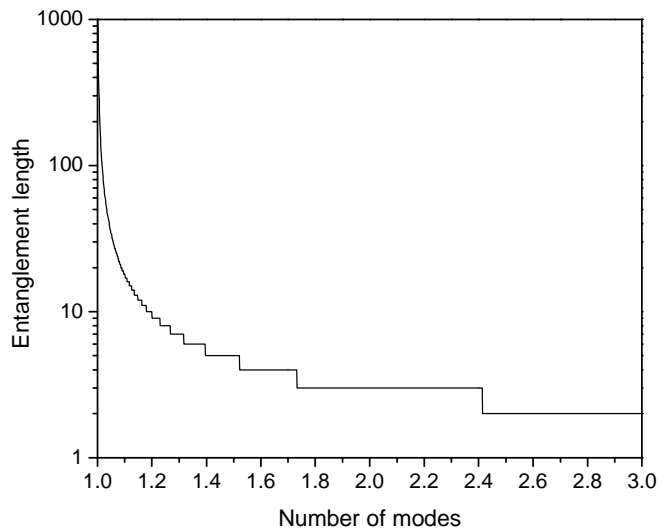}
\caption{\label{Fig2}\textbf{The entanglement length dependence on
the number of modes.} The length of the longest generated chain
where entanglement can be measured between its two end points. The
minimal length 2 is attained for more than $\sim2.41$ modes. For
chains of at least 10, 84, and 825 photons, no more than 1.2,
1.02, and 1.002 modes are allowed, respectively.}
\end{figure}

We have considered several possible causes for state imperfections
that can affect the presented scheme, such as photon partial
polarization degree, probability to emit two photons
simultaneously, imperfection of the entangling gate, and
distinguishability between the photons\cite{Orrit05}. The number
of modes that a photon occupies $N_m=\frac{1}{1-g_0^{(2)}}$ can quantify
distinguishability. The '\textit{entanglement length}' was used to
quantify the effects of various imperfections. This measure is
defined as the length of the longest linear cluster, where the
first and last photons have positive concurrence after all others
were measured at the $\{|p\rangle=\frac{1}{\sqrt{2}}(|h\rangle+|v\rangle),|m\rangle=\frac{1}{\sqrt{2}}(|h\rangle-|v\rangle)\}$ basis,
where $|h\rangle$ ($|v\rangle$) represent horizontal (vertical) polarization.
Its dependence on the number of modes is presented in Fig. \ref{Fig2}\cite{Supp}.

\begin{figure}[h]
\includegraphics[angle=0,width=86mm]{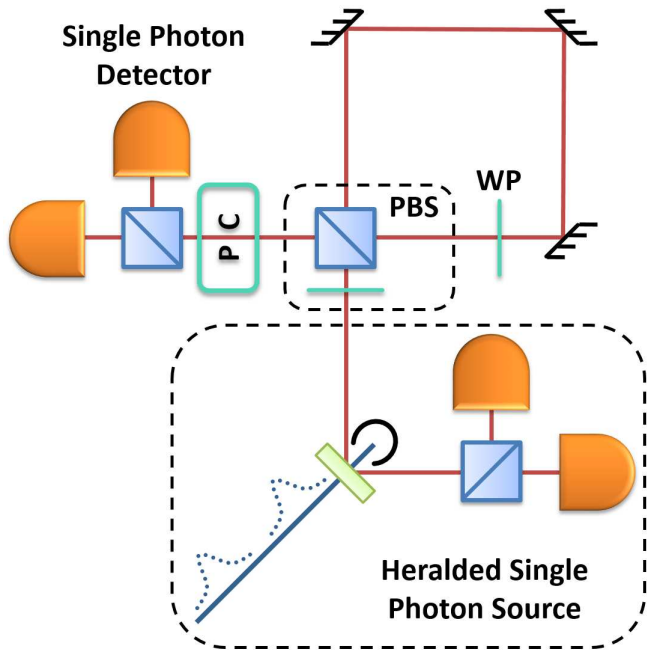}
\caption{\label{Fig3}\textbf{The optical setup used for
demonstration of the scheme.} A heralded single-photon source is
used, and the entangling operation is performed with linear optics
and post-selection. A nonlinear $\rm{BaB_2O_4}$ is pumped by a
390\,nm double Ti:Sapphire laser.}
\end{figure}

Currently, adequate on-demand single-photon sources are
unavailable. Thus, we demonstrate our scheme using a heralded
periodic PDC source. The heralded source generates two photons
probabilistically. One photon is immediately detected, announcing
the presence of its brother, which is the photon being used in the
experiment (Fig. \ref{Fig3}). Although this source is missing
determinism - a key advantage of the proposed scheme, it is still
instructive to demonstrate the basic operation of the scheme and
some of its challenges.

The entangling operation in this demonstration is by
post-selecting a single photon exiting each output port of a
PBS\cite{Pan01}. This non-unitary operation entangles photons not as
the desired CNOT gate. Nevertheless, as the source photon
polarization is known, this difference has no consequences. The
post-selection condition is fulfilled by requiring a sequential
detection of photons. Whenever two photons leave the PBS from the
same output port, there is a missing detection event. The PDC
source can simultaneously emit two pairs, or an extra pair before
or after the required sequence. Nevertheless, these are rare
higher order events with almost no effect.

By scanning the delay loop length (Fig. \ref{Fig3}), distinguishability between the
two photons is controlled. Maximal indistinguishability, and thus,
nonlocal interference, is achieved when the delay time matches the
repetition rate of the pump laser. At this length, the photons
that have already left the delay loop towards the detectors are
entangled with the last photon which is still inside the loop. In
order to detect entanglement, all photons should be observed at
the $\{|p\rangle,|m\rangle\}$ basis by a half-wave plate rotation.
The last photon which is still in the loop is observed by the
first projecting PBS and the preceding wave plate. If this photon
is detected within the consequent time slot, it was projected onto
the $|p\rangle$ state, but otherwise, its projection was onto
$|m\rangle$.

\begin{figure}[t]
\includegraphics[angle=270,width=86mm]{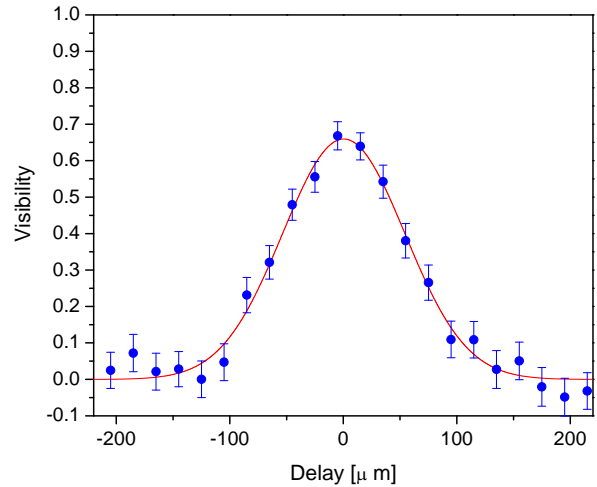}
\caption{\label{Fig4}\textbf{Nonlocal interference visibility between two
entangled photons.} A constructive (destructive) interference is
observed as the delay is scanned, for the $|\phi^+\rangle$
($|\phi^-\rangle$) state. The fourfold signal is composed of the
detection of the two heralding photons and the two entangled
photons. Each data point was integrated over 16\,minutes.
Event rate was of 1 $Hz$ at an average pump power of 350\,$mW$.
Errors are calculated assuming Poisson distribution.}
\end{figure}

\begin{figure}[t]
\includegraphics[angle=270,width=86mm]{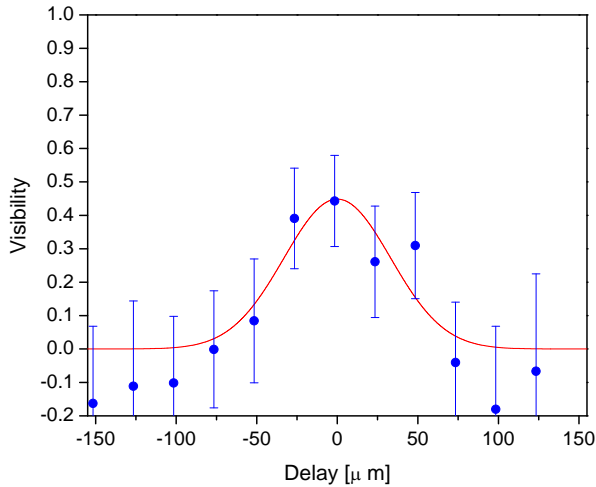}
\caption{\label{Fig5}\textbf{Nonlocal interference visibility between three
entangled photons.} A constructive (destructive) interference is
observed as the delay is scanned, for the $\frac{1}{\sqrt{2}}(|\phi^+h\rangle+|\phi^-v\rangle)$
($\frac{1}{\sqrt{2}}(|\phi^+h\rangle-|\phi^-v\rangle)$) state. The sixfold signal is composed of the
detection of the three heralding photons and the three entangled
photons. Each data point was integrated over 34\,hours.
Event rate was of 1 per 35\,minutes at an average pump power of 550\,$mW$.
Errors are calculated assuming Poisson distribution.}
\end{figure}

The interference graphs of two and three photon states are
presented in Figs. \ref{Fig4} and \ref{Fig5}, respectively. When
distinguishability is removed, some polarization sequences
interfere constructively and some destructively. The two photon
state interference visibility is $V_2=66\pm2\%$ and the three
photon state visibility is $V_3=45\pm11\%$. We estimate that
higher order events account for 5-10\% of the two photon
visibility degradation, as well as imperfect PBS performance.
Nevertheless, spectral distinguishability is probably the major
cause. Thus, it is possible to use $V_2$ to estimate the number of
distinguishable modes to be $N_m=\frac{1}{V_2}\sim1.5$,
corresponding to an expected three photon visibility of
$V_3=V_2^2=44\%$, and maximum possible entanglement length of 5.

It is instructive to examine the prospects of currently available
sources. Recently, high levels of indistinguishability have been
demonstrated\cite{Wei14,Somaschi15}, corresponding to possible
entanglement lengths of 80 and 150 photons. The second work
reports also 0.65 collected photons per pump pulse, suggesting the
detection of a 32 photon event every second for the 82\,MHz
pumping rate. In practice, the overall collection and detection
efficiencies of these two works are below 1\%, limiting the
possible detection rate to around 3-4 photon events every second.
However, this is in principle a technical limitation that can be
improved by better optics and detectors. Overall detection
efficiency of about ~5\% has already been achieved\cite{Strauff07,Loredo16},
enabling a sixfold event detection
every second. Additionally, considering almost perfect detection
efficiency\cite{Lita08,Pernice12}, a 9 photon event can be
detected every second, and 12 every 6 minutes. The short lifetime
of single-photon emitters compared to the repetition rate enables
photon sources that are pumped by faster lasers\cite{Bartels08}.

In conclusion, we present a compact time-multiplexed scheme for
generating multi-photon linear cluster states, which is realizable
with current single-photon sources. It holds promise for better
scalability and less need for tunability, and thus the possibility
to achieve much larger entangled multi-photon states than it is
possible today. A crucial component is a photon-photon entangling
gate. Currently, the most efficient gate is obtained by linear optics
and post-selection, but there are many efforts pursuing exactly
this goal. Finally, we should mention the possibility to append
delay loops of different lengths. Such combinations can generate
cluster states of higher dimensionality, and even graph states
which are useful for the one-way quantum computer scheme. We will
describe these options in a future publication.

\subsection{Gate operation.}
Following Fig. \ref{Fig1}, a photon is set at the horizontally
polarized state $|h\rangle$ at position 1.
The photon
reaches a CNOT entangling gate, but passes it through with no
effect into the delay loop at position 2. When the photon reaches
position 3, it is at the $|p\rangle$ state, after passing a
half-wave plate that rotates the polarization by $45^\circ$.

Synchronized with the second arrival of this photon to the
entangling gate, the single photon source is triggered again such
that the two photons are simultaneously entering both input ports
of the entangling gate at positions 1 and 3. Photons 1 and 3 are
the target and control qubits, respectively. The CNOT gate
operation\cite{Supp} entangles the following states
\begin{eqnarray}\label{CNOT}
\nonumber
|h_1p_3\rangle&\rightarrow&|\phi^+_{24}\rangle=\frac{1}{\sqrt{2}}(|h_2h_4\rangle+|v_2v_4\rangle)\\
\nonumber
|h_1m_3\rangle&\rightarrow&|\phi^-_{24}\rangle=\frac{1}{\sqrt{2}}(|h_2h_4\rangle-|v_2v_4\rangle)\\
\nonumber
|v_1p_3\rangle&\rightarrow&|\psi^+_{24}\rangle=\frac{1}{\sqrt{2}}(|v_2h_4\rangle+|h_2v_4\rangle)\\
|v_1m_3\rangle&\rightarrow&|\psi^-_{24}\rangle=\frac{1}{\sqrt{2}}(|v_2h_4\rangle-|h_2v_4\rangle)\,,
\end{eqnarray}
thus, the $|h_1p_3\rangle$ state becomes the $|\phi^+_{24}\rangle$
Bell state (indices indicate the photon positions), and
entanglement is created between the photon that has left the loop
and the other one that is still inside it. After the photon inside
the loop is rotated by the wave plate, the source is synchronously
triggered for the third time, and the three photon state is
$\frac{1}{\sqrt{2}}|h_1\rangle\otimes(|p_3h_5\rangle+|m_3v_5\rangle)$.
This time, the entangling operation creates the three photon
entangled state
$\frac{1}{\sqrt{2}}(|\phi^+_{24}h_5\rangle+|\phi^-_{24}v_5\rangle)$.
This is a three photon GHZ state\cite{Greenberger90}, as it can
also be written as
$\frac{1}{\sqrt{2}}(|h_2h_4p_5\rangle+|v_2v_4m_5\rangle)$.

In order to see the results of adding photons continuously in a
similar way, we examine just one more stage. When a fourth photon
is added after the photon polarization inside the loop is rotated,
the state becomes
$\frac{1}{\sqrt{2}}|h_1\rangle\otimes(|p_3h_5p_6\rangle+|m_3v_5m_6\rangle)$.
The entangling operation creates a four photon entangled state
$\frac{1}{\sqrt{2}}(|\phi^+_{24}h_5p_6\rangle+|\phi^-_{24}v_5m_6\rangle)$,
which is actually the linear cluster state
$\frac{1}{2}(|h_2h_4h_5p_6\rangle+|v_2v_4h_5p_6\rangle+|h_2h_4v_5m_6\rangle-|v_2v_4v_5m_6\rangle)$.
In graph state language, also the two photon Bell state and the
three photon GHZ states are linear cluster states. Thus, every
additional photon is entangled to the end of the current linear
cluster state, extending it by one unit. Interestingly, if the
wave-plate in the delay loop is pulled out after the first photon
has already passed it, The continuously growing state is of the
GHZ type \cite{Greenberger90}. This pulling out can be achieved be
realizing the wave-plate with a Pockels cell.

\subsection{Experimental PDC setup.}
The heralded source generates two
photons probabilistically with the PDC process.
As we use a PDC source for polarization entangled photons, the
generated photons are of horizontal or vertical polarizations
randomly. Respectively, we collect data for any sequence
combination of input photon polarizations.

The entangling operation in this demonstration is using linear
optical elements and post-selection \cite{Pan01}. The optical
elements are a PBS with a half-wave plate at $22.5^\circ$ before
one of the input ports. When combined with the post-selection of a
single photon coming out of each of the two output ports, this
combination entangles both the $|h_1p_3\rangle$ and the
$|v_1m_3\rangle$ states into the same $|\phi_{24}^+\rangle$ state,
and the $|v_1p_3\rangle$ and the $|h_1m_3\rangle$ states into the
$|\phi_{24}^-\rangle$ state.

Maximal indistinguishability, and thus, nonlocal interference, is achieved
when the delay time matches the repetition rate of the pump laser.
At this length, one photon that has left the delay loop towards
the detectors is entangled with the other photon which is still
inside the loop. As the first photon which was emitted by the
single-photon source was required to pass the PBS into the loop,
it was projected onto the $|h\rangle$ state. Thus, this photon's
original polarization has no consequences. The polarization of the
second emitted photon though, determines whether the final state
will be $|\phi^+\rangle$ or $|\phi^-\rangle$. In order to detect
these two states, the polarization of both photons should be
observed at the $\{|p\rangle,|m\rangle\}$ basis. The first photon
that has left the loop is thus rotated by a half-wave plate and
detected beyond a second PBS. The polarization state of the other
photon which is still in the loop is observed differently.
Luckily, the wave plate which is already inside the loop rotates
it to the right basis. Its measurement occurs at the first
projecting PBS. If this photon is projected onto the $|h\rangle$
state that corresponds to $|p\rangle$ before the wave plate, it
leaves the loop. As there is another wave plate before the second
PBS, it will be detected at the next time slot by any one of the
two detectors, with equal probabilities. On the other hand, if the
second photon is projected onto the $|v\rangle$ state, it will be
delayed in the loop for another round, after which it will have
equal chances of leaving the loop or staying for another round.
Overall, if the second photon is detected in the time slot right
after the first one, it was projected onto the $|h\rangle$ state,
but if it is detected on any later time, its projection was onto
$|v\rangle$.

The three photon interference was observed at the
$\{|p\rangle,|m\rangle\}$ basis in a similar way. Additional
birefringent phase of $90^\circ$ was introduced inside the delay
loop. See Supplementary Material for further
explanations\cite{Supp}.


\textbf{Acknowledgements} The authors would like to thank the Israeli Science
Foundation for supporting this work under grants 546/10 and
793/13.

\end{document}



\raggedbottom
\section*{Supplementary information}

\section{Effects of imperfections}

In what follows we present a detailed analysis of the performance
of our scheme under the effect of the main sources of error. These
include errors in the initial polarization state of the incoming photons,
imperfection of the polarized beam splitter used within the setup,
non-ideal indistinguishability of photons, and the simultaneous emission
of two photons (instead of one photon) by the single-photon source.
Specifically, we show how any of these errors affects the performance
of the scheme dependent on the number of photons and on the severity
of the imperfection. Our analysis considers the effect of each error
source separately. Nevertheless, this provides us enough insight in
order to understand which are the main limiting factors of our scheme
and what is the expected efficiency of the implementation of our scheme
using state of the art single-photon sources. We use three figures
of merits in order to quantify the performance of the scheme. The
first figure of merit, which we denote by $\mathcal{C}_{1,N}$, is
the concurrence of the final two-qubit state of the first and last
qubits, which is obtained after measuring all other qubits along the
$\hat{y}$ axis. The second figure of merit, which we denote by $\mathcal{L}$,
characteristics the ``\emph{entanglement length}'' of the obtained
$N-$qubit linear cluster state, and is defined as the length of the longest
linear cluster state for which $\mathcal{C}_{1,N}$ is still positive.
The third figure of merit, which is denoted by $\mathcal{F}$, is
the fidelity between the obtained $N-$qubit cluster state and the
\emph{ideal} $N-$qubit cluster state. Formally, the three figures
of merits are defined by
\begin{eqnarray}
\mathcal{C}_{1,N} & = & \mathcal{C}\left(\mathrm{Tr}{}_{2,...,N-1}\left(\frac{\Pi_{y}\rho\Pi_{y}}{\mathrm{Tr}\left(\Pi_{y}\rho\right)}\right)\right),\\
\mathcal{L} & = & \max\, N\quad\mathrm{s.t}.\quad\mathcal{C}_{1,N}>0,\\
\mathcal{F} & = & \left\langle \psi_{ideal}\right|\rho\left|\psi_{ideal}\right\rangle ,
\end{eqnarray}
 where $\Pi_{y}=\mathbb{I}_{1}\otimes\prod_{i=2}^{N-1}\left|\uparrow_{y}^{i}\right\rangle \left\langle \uparrow_{y}^{i}\right|\otimes\mathbb{I}_{N}$,
$\mathcal{C}$ is the concurrence, and $\left|\psi_{ideal}\right\rangle $
is the ideal $N-$qubit cluster state.

\subsection{Distinguishability of photons}

The desired interference between two post-selected photons can be
obtained with certainty only when the two photons are identical in
all of their non-polarization degrees of freedom. These include their
frequency, spatial and temporal modes. When the photons differ in some of their non-polarization degrees of freedom the probability
of interference is decided by the mean magnitude-squared overlap between
the wave-functions of the photons, which quantifies the photons` ``\emph{probability
of indistinguishability}''.

Consider a single photon in a pure state, which can be described by
\[
\left|\omega_{i}\right\rangle =\int_{-\infty}^{+\infty}d\upsilon\phi_{\omega_{i}}\left(\upsilon\right)\left|\upsilon\right\rangle \text{,}
\]
where $\phi_{\omega_{i}}\left(\upsilon\right)$ is a normalized spectral
amplitude function, such that $\int_{-\infty}^{+\infty}d\upsilon\left|\phi_{\omega_{i}}\left(\upsilon\right)\right|^{2}=1$.
The probability of indistinguishability of two independent such photons
is given by their squared overlap, $\mathcal{I}_{i,j}=\left|\left\langle \omega_{i}|\omega_{j}\right\rangle \right|^{2}.$
Similarly, for photons in a mixed state,
\[
\rho_{i}=\int_{-\infty}^{+\infty}d\omega_{i}f\left(\omega_{i}\right)\left|\omega_{i}\right\rangle \left\langle \omega_{i}\right|,
\]
 where $f\left(\omega_{i}\right)$ is a normalized probability distribution,
such that $\int_{-\infty}^{+\infty}d\omega_{i}f\left(\omega_{i}\right)=1$,
the probability of indistinguishability can be defined by their mean
magnitude-squared overlap \cite{Sun},
\[
\mathcal{I}=\mathrm{Tr}\left(\rho_{i}\rho_{j}\right)=\iint_{-\infty}^{+\infty}d\omega_{i}d\omega_{j}f\left(\omega_{i}\right)g\left(\omega_{j}\right)\left|\left\langle \omega_{i}|\omega_{j}\right\rangle \right|^{2}.
\]
 Indeed, the probability of indistinguishability coincides with the
interference visibility of the Hong-Ou-Mandel (HOM) dip. In an HOM experiment, the
coincidence probability of two independent photons, $\rho_{i}$ and
$\rho_{j}$, is given by
\[
P_{cc}\left(\delta\tau\right)=\frac{1}{2}\left(1-\mathcal{I}\left(\delta\tau\right)\right),
\]
 where $\mathcal{I}\left(\delta\tau\right)=\iiiint_{-\infty}^{+\infty}d\omega_{i}d\omega_{j}d\upsilon_{1}d\upsilon_{2}f\left(\omega_{i}\right)g\left(\omega_{j}\right)\phi_{\omega_{i}}^{*}\left(\upsilon_{1}\right)\phi_{\omega_{j}}\left(\upsilon_{1}\right)\phi_{\omega_{j}}^{*}\left(\upsilon_{2}\right)\phi_{\omega_{i}}\left(\upsilon_{2}\right)e^{-i\delta\tau\left(\upsilon_{2}-\upsilon_{1}\right)}$
is the probability of indistinguishability of the two photons with
a time delay of $\delta\tau$ between the arrival time of the photons
at the beam splitter. In terms of the initial states of the photons,
the visibility is then given by \cite{Ou}
\[
V=\frac{\max_{\delta\tau}P_{cc}\left(\delta\tau\right)-\min_{\delta\tau}P_{cc}\left(\delta\tau\right)}{\max_{\delta\tau}P_{cc}\left(\delta\tau\right)}=\frac{P_{cc}\left(\infty\right)-P_{cc}\left(0\right)}{P_{cc}\left(\infty\right)}=\mathcal{I}\left(0\right)=\mathcal{I}.
\]
 Note that as $V=\mathcal{I}=\mathrm{Tr}\left(\rho_{i}\rho_{j}\right)=\frac{\mathrm{Tr}\left(\rho_{i}^{2}\right)+\mathrm{Tr}\left(\rho_{j}^{2}\right)-2\left\Vert \rho_{i}-\rho_{j}\right\Vert ^{2}}{2}$,
where $\left\Vert \rho\right\Vert ^{2}=\mathrm{Tr}\left(\rho^{\dagger}\rho\right)$,
it is clearly seen that the probability of indistinguishability depends
on both the purity of the states of the photons and their identicality.

In our experimental setup we considered polarization photonic states,
which are entangled by the operation of a PBS and post-selection.
Following Mandel \cite{Mandel}, we will now show that in our setup,
as in an HOM experiment, the probability of indistinguishability is
related to normalized coincidence probability by
\[
\mathcal{I}=1-g_{p,m}^{2}\left(0\right),
\]

where $g_{p,m}^{2}\left(\delta\tau\right)$ is obtained from the normalized
second-order correlation function by
\[
g_{p,m}^{2}\left(\delta\tau\right)=\iint_{-\infty}^{+\infty}dtd\tau\frac{\left\langle E_{p,3}^{-}\left(t\right)E_{m,4}^{-}\left(t+\tau\right)E_{m,4}^{+}\left(t+\tau\right)E_{p,3}^{+}\left(t\right)\right\rangle }{\left\langle E_{p,3}^{-}\left(t\right)E_{p,3}^{+}\left(t\right)\right\rangle \left\langle E_{p,4}^{-}\left(t+\tau\right)E_{p,4}^{+}\left(t+\tau\right)\right\rangle },
\]
and $E_{i,j}^{\pm}$ are the field operators corresponding to a photon
with a polarization $i$ at the location $j$. We note that for the
$\left|p_{1}p_{2}\right\rangle $ input state, the resulting post-selected
state of two indistinguishable photons is the maximally entangled
state $\left|\phi^+_{id}\right\rangle =\frac{1}{\sqrt{2}}\left(\left|h_{3}h_{4}\right\rangle +\left|v_{3}v_{4}\right\rangle \right)$,
and we thus denote its density matrix,
\[
\rho_{id}=\frac{1}{2}\begin{pmatrix}1 & 0 & 0 & 1\\
0 & 0 & 0 & 0\\
0 & 0 & 0 & 0\\
1 & 0 & 0 & 1
\end{pmatrix},
\]
 as the indistinguishable density matrix, where the rows correspond
to the states $\left|p_{3}p_{4}\right\rangle $, $\left|p_{3}m_{4}\right\rangle $,
$\left|m_{3}p_{4}\right\rangle $, and $\left|m_{3}m_{4}\right\rangle $.
Similarly, the resulting state of two fully distinguishable photons
is given by the mixed state
\[
\rho_{d}=\frac{1}{4}\begin{pmatrix}1 & 0 & 0 & 1\\
0 & 1 & 1 & 0\\
0 & 1 & 1 & 0\\
1 & 0 & 0 & 1
\end{pmatrix}.
\]
 A general resulting state,
\[
\rho=\begin{pmatrix}\rho_{11} & 0 & 0 & \rho_{14}\\
0 & \rho_{22} & \rho_{23} & 0\\
0 & \rho_{32} & \rho_{33} & 0\\
\rho_{41} & 0 & 0 & \rho_{44}
\end{pmatrix},
\]
can therefore be decomposed as
\[
\rho=p_{id}\rho_{id}+p_{d}\rho_{d},
\]
where $p_{id}=\mathcal{I}$ is the probability of indistinguishability,
and $p_{id}+p_{d}=1$. From this we conclude that
\[
p_{id}=\rho_{11}+\rho_{44}-\rho_{22}-\rho_{33}.
\]
The calculation of the normalized coincidence probability yields
\[
g_{p,m}^{2}\left(0\right)=\frac{\mathrm{Tr}\left(a_{p,3}^{\dagger}a_{m,4}^{\dagger}a_{m,4}a_{p,3}\rho\right)}{\mathrm{Tr}\left(a_{p,3}^{\dagger}a_{p,3}\rho\right)\mathrm{Tr}\left(a_{m,4}^{\dagger}a_{m,4}\rho\right)}=4\rho_{22}=p_{d},
\]
and hence %
\footnote{We actually calculate $g_{p,m}^{2}\left(\delta\tau\right)$, where
the effect of a time delay enters via the state $\rho$. We assume,
however, that in our scenario $\delta\tau=0$ and that the indistinguishability
is decided only by the source of the photons%
},
\[
\mathcal{I}=1-g_{p,m}^{2}\left(0\right).
\]
 For example, in the case of a two-level system (TLS) single-photon source it can be
shown that $g^{2}\left(0\right)=1-\frac{T_{2}}{2T_{1}},$ so $\mathcal{I}=\frac{T_{2}}{2T_{1}}$
\cite{Bylander}. In the case of heralded single-photons emitted by
a SPDC setup, the probability of indistinguishability is given by
$\mathcal{I}=\frac{\sigma_{int}}{\sigma},$ where $\sigma=\sqrt{\sigma_{int}^{2}+\sigma_{ext}^{2}}$
, $\sigma_{int}$ is the (transform-limited) Gaussian width of the
photon's spectrum (intrinsic width), and $\sigma_{ext}$ is the Gaussian
width of the spectrum of the center frequency, resulting by tracing
out the state of the detected twin photon (extrinsic width) \cite{Sun}.

In terms of the post-selected state, the visibility (in the $\left\{ p,m\right\} $
basis) is given by
\[
V_{p.s}=\mathrm{Tr}\left(\sigma_{z}\otimes\sigma_{z}\rho\right)=\rho_{11}+\rho_{44}-\rho_{22}-\rho_{33}=\mathcal{I}
\]
and indeed, by the definition of $V_{p.s}=\frac{P_{same}-P_{opp}}{P_{same}+P_{opp}}$,
where $P_{same}$ $\left(P_{opp}\right)$ denote the probability that
the two photons arrive at the same (different) polarization, we have that
$V_{p.s}=P_{same}^{max}-P_{opp}^{min}=1-2P_{opp}^{min}=1-\left(1-\mathcal{I}\left(0\right)\right)=\mathcal{I}$.

\begin{figure}[H]
\centering{}\includegraphics[scale=0.78]{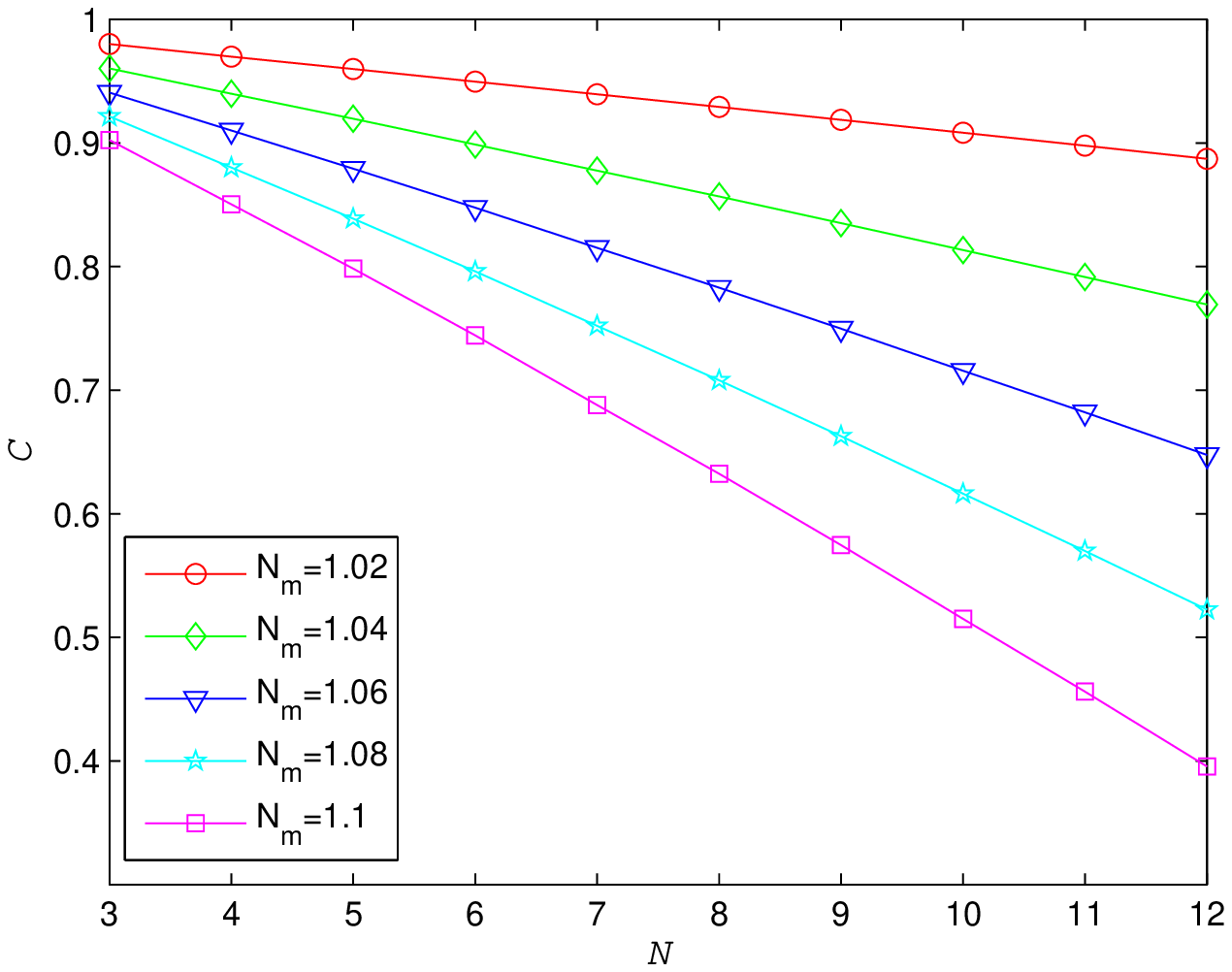}\protect\caption{$\mathcal{C}$ as a function of the number of photons.}
\label{Fig6}
\end{figure}

\begin{figure}[H]
\centering{}\includegraphics[scale=0.78]{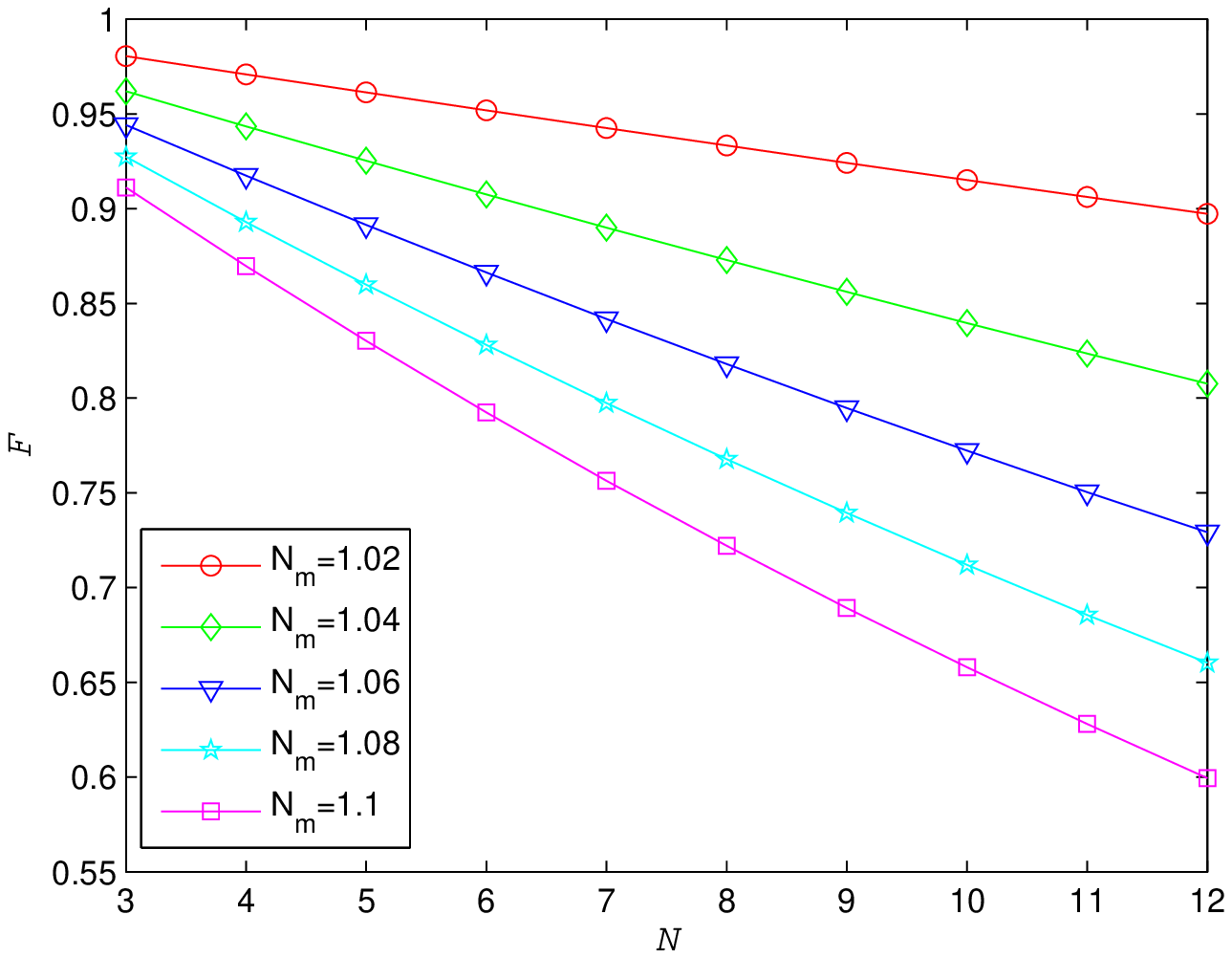}\protect\caption{$\mathcal{F}$ as a function of the number of photons. }
\label{Fig7}
\end{figure}

\begin{figure}[H]
\centering{}\includegraphics[scale=0.78]{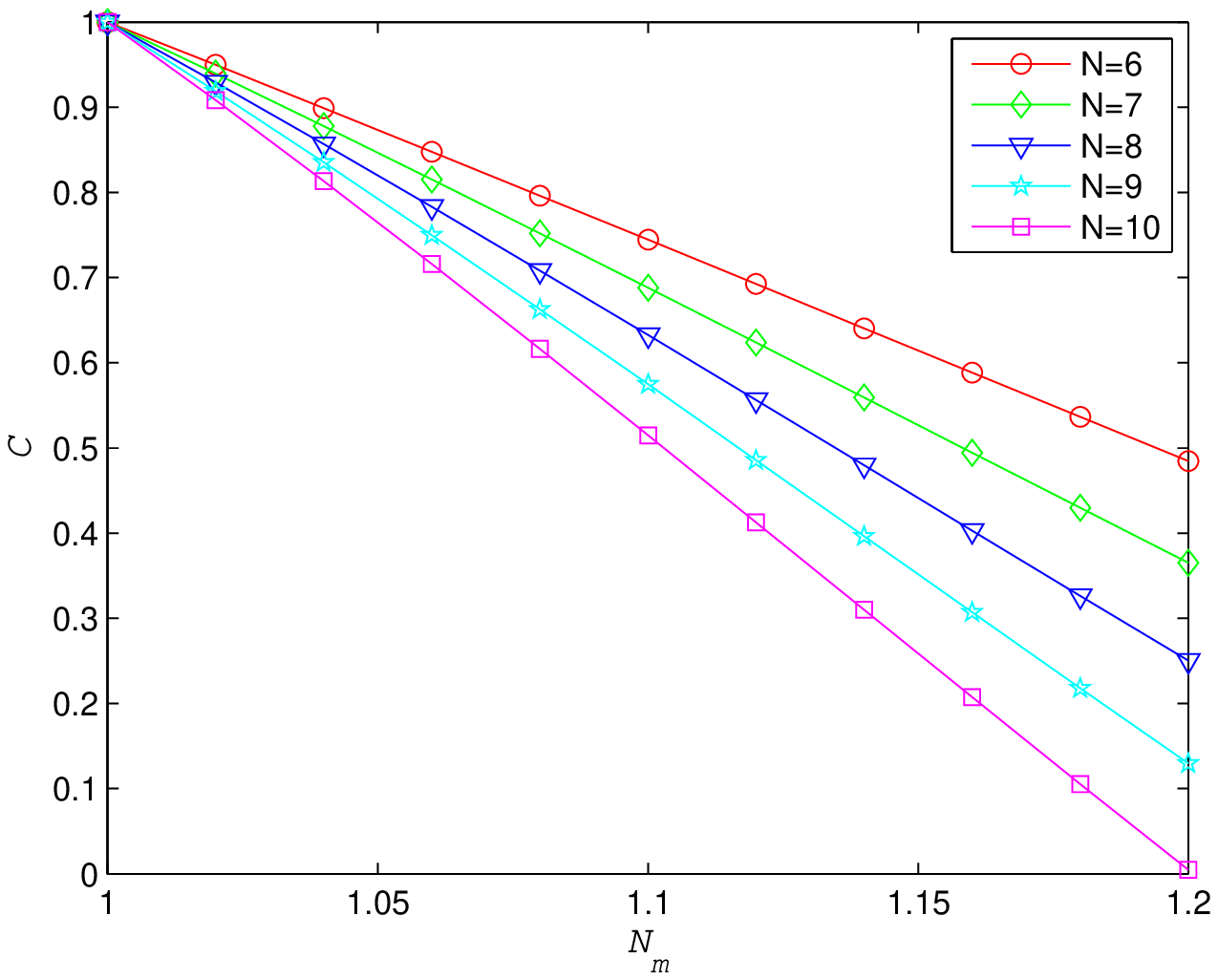}\protect\caption{$\mathcal{C}$ as a function of the number of modes. }
\label{Fig8}
\end{figure}

\begin{figure}[H]
\centering{}\includegraphics[scale=0.78]{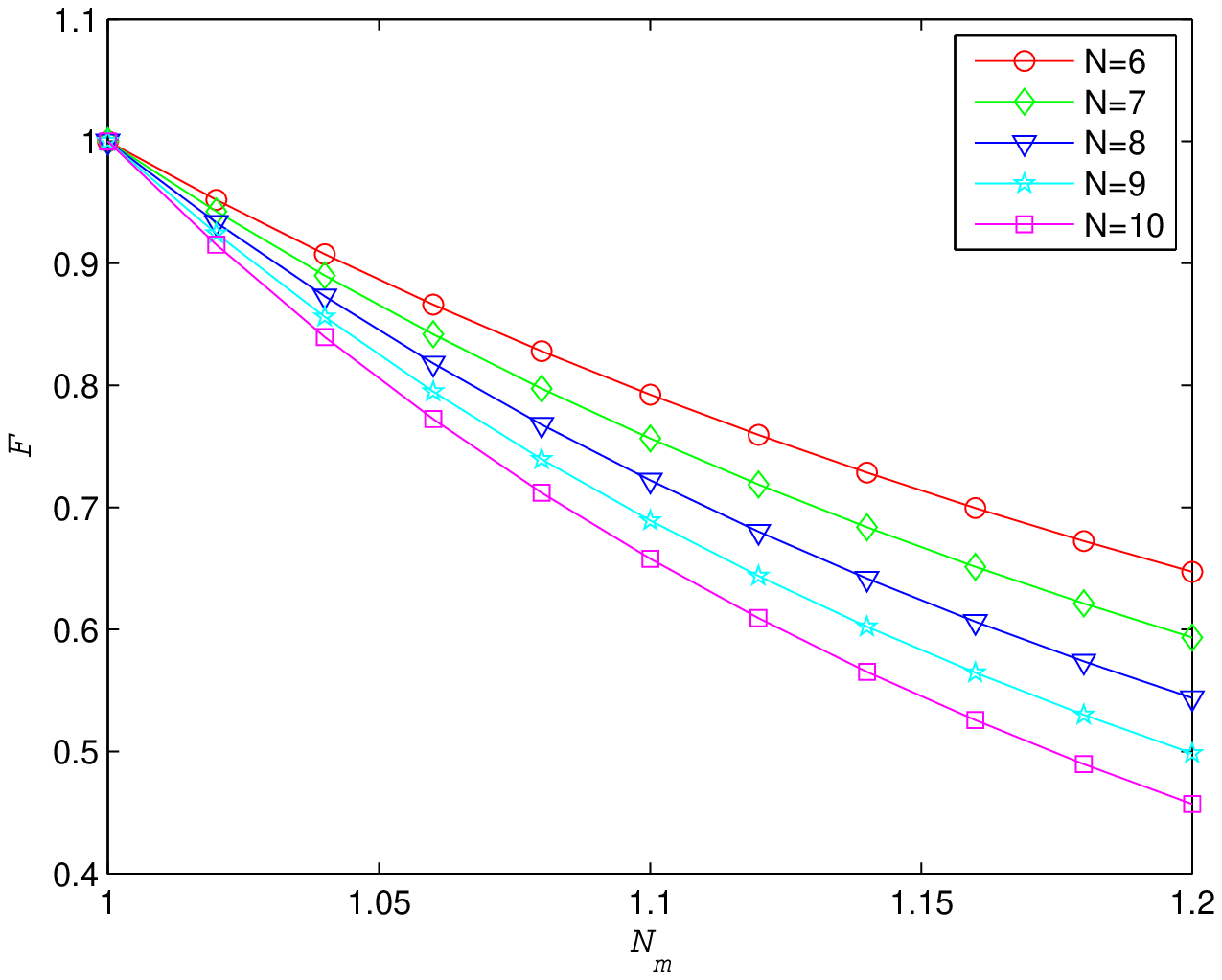}\protect\caption{$\mathcal{F}$ as a function of the number of modes. }
\label{Fig9}
\end{figure}

The probability of indistinguishability can also be interpreted by
modeling the photon's state (of all of its other degrees of freedom)
as a mixd state of uniformly distributed (orthogonal) modes. In this
case $\mathcal{I}$ is just the probability that the two photons have
the same mode, and hence, the ``\emph{number of modes}'' is given
by
\[
N_{m}=\frac{1}{\mathcal{I}},
\]
which also implies the relation
\[
g_{p,m}^{2}\left(0\right)=\frac{N_{m}-1}{N_{m}}.
\]
In this picture, the post-selected process of the PBS can be described
by
\[
\varepsilon\left(\rho\right)=\frac{1}{N_{m}}\varepsilon_{0}\rho\varepsilon_{0}+\frac{N_{m}-1}{2N_{m}}\left(\varepsilon_{0}\rho\varepsilon_{0}+\varepsilon_{1}\rho\varepsilon_{1}\right),
\]
where $\varepsilon_{0}=\sigma_{0}\otimes\sigma_{z}+\sigma_{z}\otimes\sigma_{0}$
describes the desired process, and $\varepsilon_{1}=\sigma_{0}\otimes\sigma_{0}+\sigma_{z}\otimes\sigma_{z}$,
which means that the process $\frac{1}{2}\left(\varepsilon_{0}\rho\varepsilon_{0}+\varepsilon_{1}\rho\varepsilon_{1}\right)$
takes the input polarization state $\left|p\right\rangle _{1}\left|p\right\rangle _{2}$
to the (equally) \emph{mixed} state of the output states $\left|h\right\rangle _{3}\left|h\right\rangle _{4}$
and $\left|v\right\rangle _{3}\left|v\right\rangle _{4}$. In Fig.
\ref{Fig6} and Fig. \ref{Fig7} we plot $\mathcal{C}$ and $\mathcal{F}$ as a function
of the number of photons for a fixed number of modes, while in Fig.
\ref{Fig8} and Fig. \ref{Fig9} we plot $\mathcal{C}$ and $\mathcal{F}$ as a function
of the number of modes for a fixed number of photons. $\mathcal{L}$
as function of $N_{m}$ is shown in Fig. 2 in the main text.

\subsection{Polarization errors}

\begin{figure}[H]
\centering{}\includegraphics[scale=0.78]{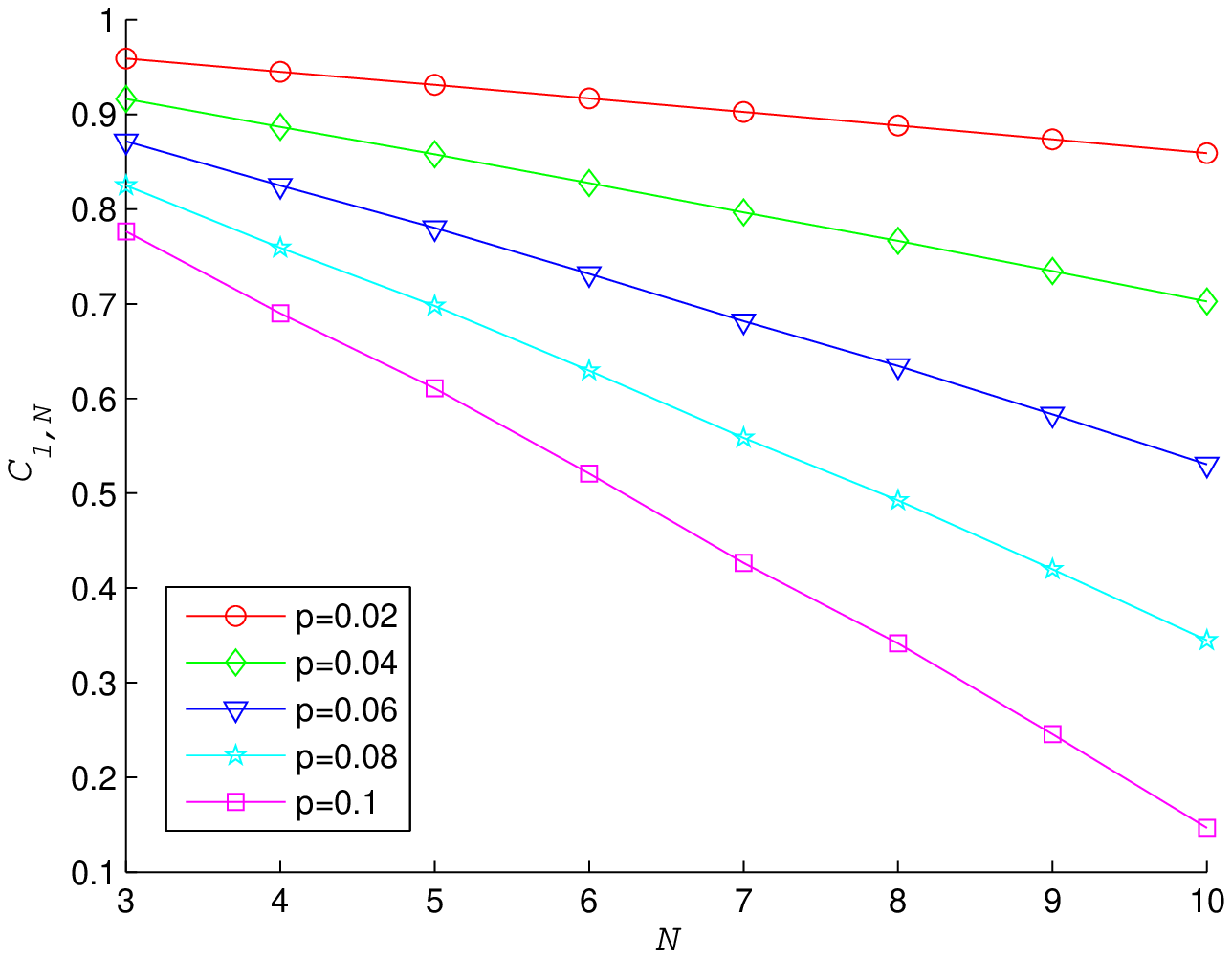}\protect\caption{$\mathcal{\mathcal{C}}_{1,N}$ as a function of the number of qubits
for different values of $p$ (depolarizing channel). }
%
%
\label{Fig1}
\end{figure}

\begin{figure}[H]
\centering{}\includegraphics[scale=0.78]{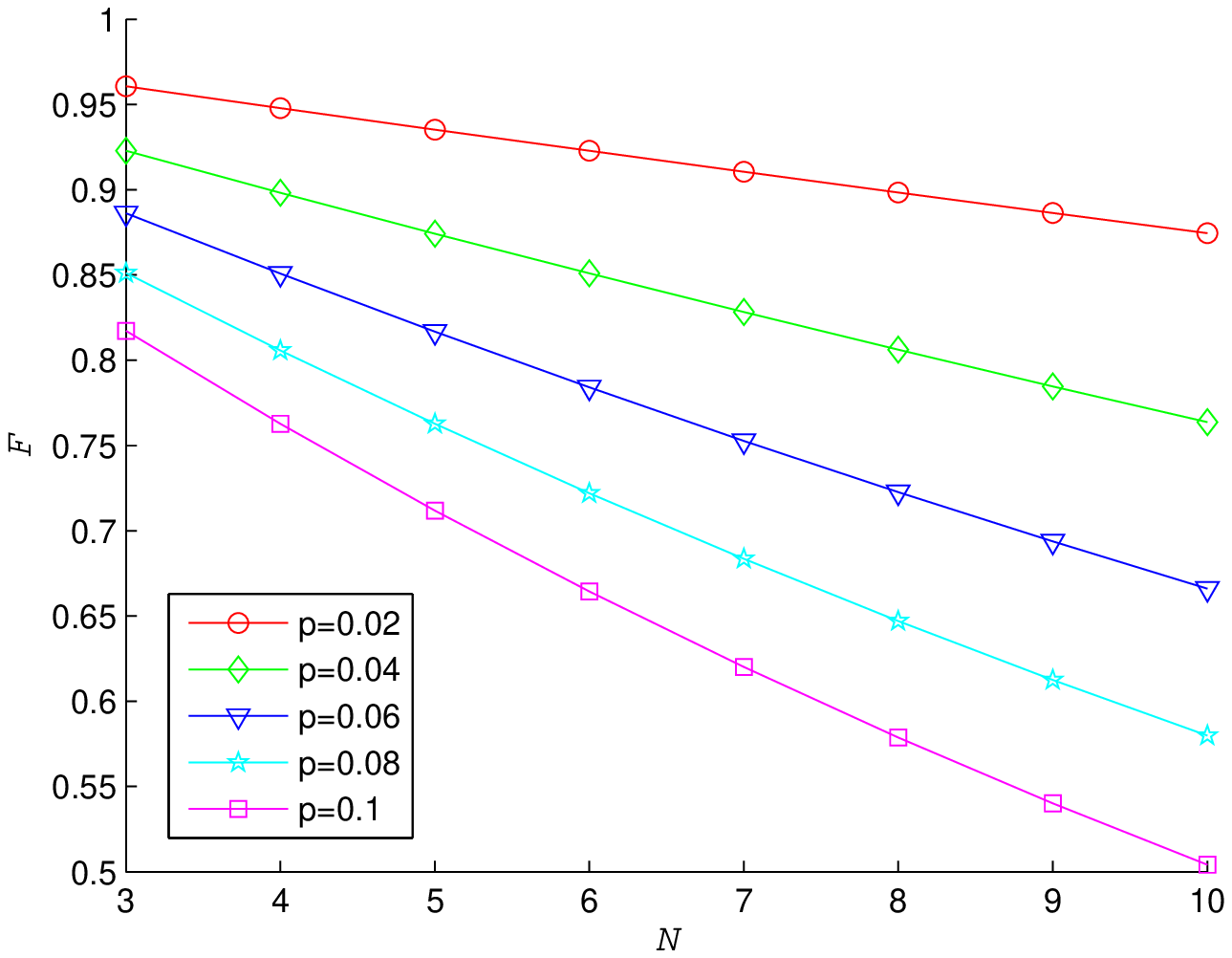}\protect\caption{$\mathcal{F}$ as a function of the number of qubits for different
values of $p$ (depolarizing channel). }
%
%
\label{Fig2}
\end{figure}

\begin{figure}[H]
\centering{}\includegraphics[scale=0.78]{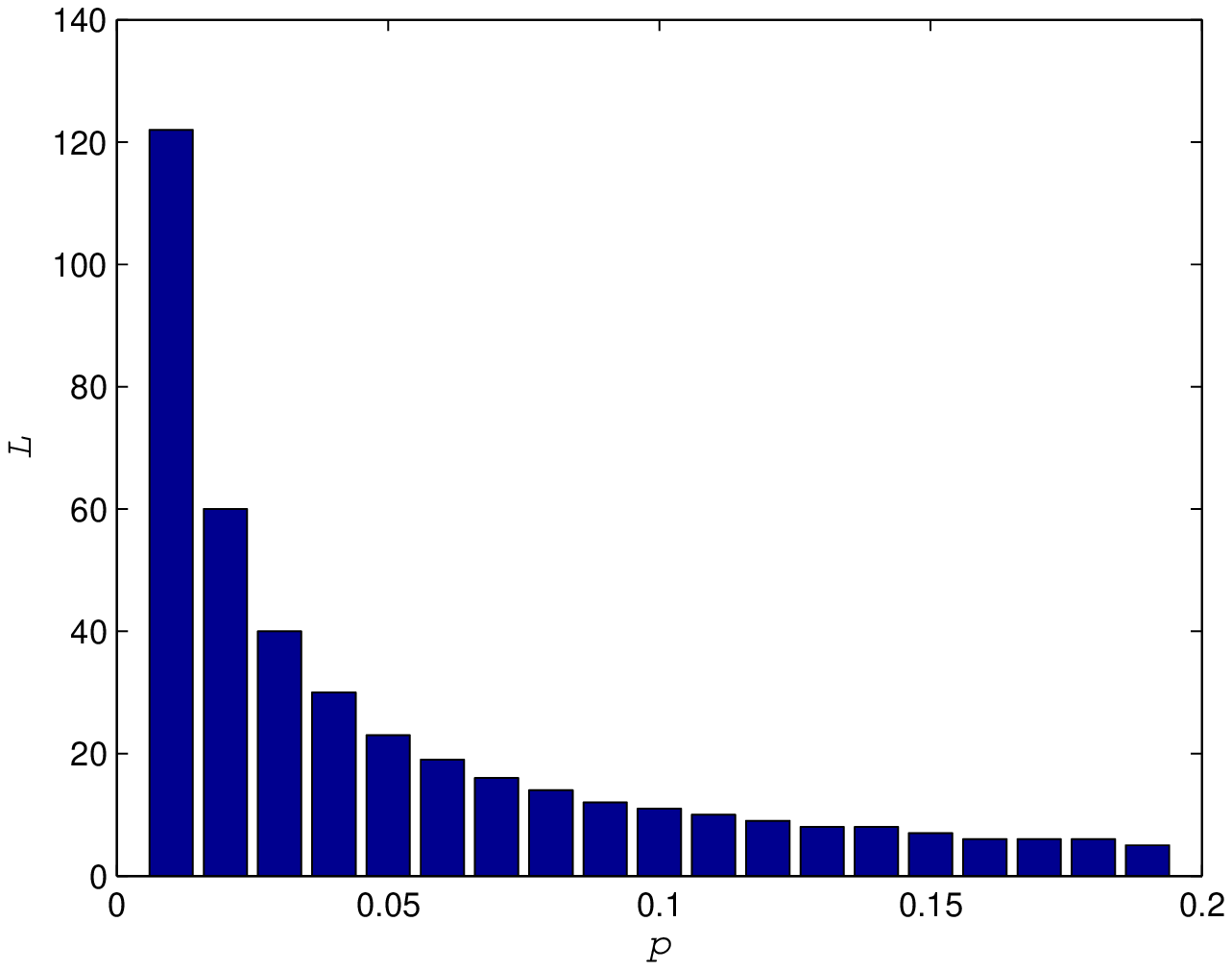}\protect\caption{$\mathcal{L}$ as a function of $p$ (depolarizing channel). }
%
%
\label{Fig3}
\end{figure}

Ideally, the initial polarization state of each photon which enters
the setup is a $\left|p\right\rangle =\frac{1}{\sqrt{2}}\left(\left|h\right\rangle +\left|v\right\rangle \right)$
state ($|h\rangle$ ($|v\rangle$) represent horizontal (vertical) polarization and for further use
we define $|m\rangle=\frac{1}{\sqrt{2}}(|h\rangle-|v\rangle)\}$). However, errors in the preparation will result in imperfect
polarization states. These errors are usually caused by the presence
of entanglement between a photon's polarization and its other degrees
of freedom, or an entanglement between a photon and its environment
(as is the case in photons emitted by parametric down-conversion).
We model these polarization errors by a depolarizing quantum channel
\cite{Nielsen},
\[
\varepsilon\left(\rho\right)=\left(1-p\right)\rho+\frac{p}{3}\left(\sigma_{x}\rho\sigma_{x}+\sigma_{y}\rho\sigma_{y}+\sigma_{z}\rho\sigma_{z}\right),
\]
$\rho$ is the density matrix, and assume that each photon undergoes a depolarizing channel before
it enters the setup. $\mathcal{C}_{1,N}$ and $\mathcal{F}$ as function
of the number of qubits for different values of $p$ are shown in
Fig. \ref{Fig1} and Fig. \ref{Fig2} respectively. $\mathcal{L}$ as function of $p$
is shown in Fig. \ref{Fig3}.

Polarization errors can be compensated by adding a polarization filter.
Of course, this will result in a better accuracy but with a lower
efficiency.

\subsection{Imperfect polarization beam splitter}

\begin{figure}[H]
\centering{}\includegraphics[scale=0.78]{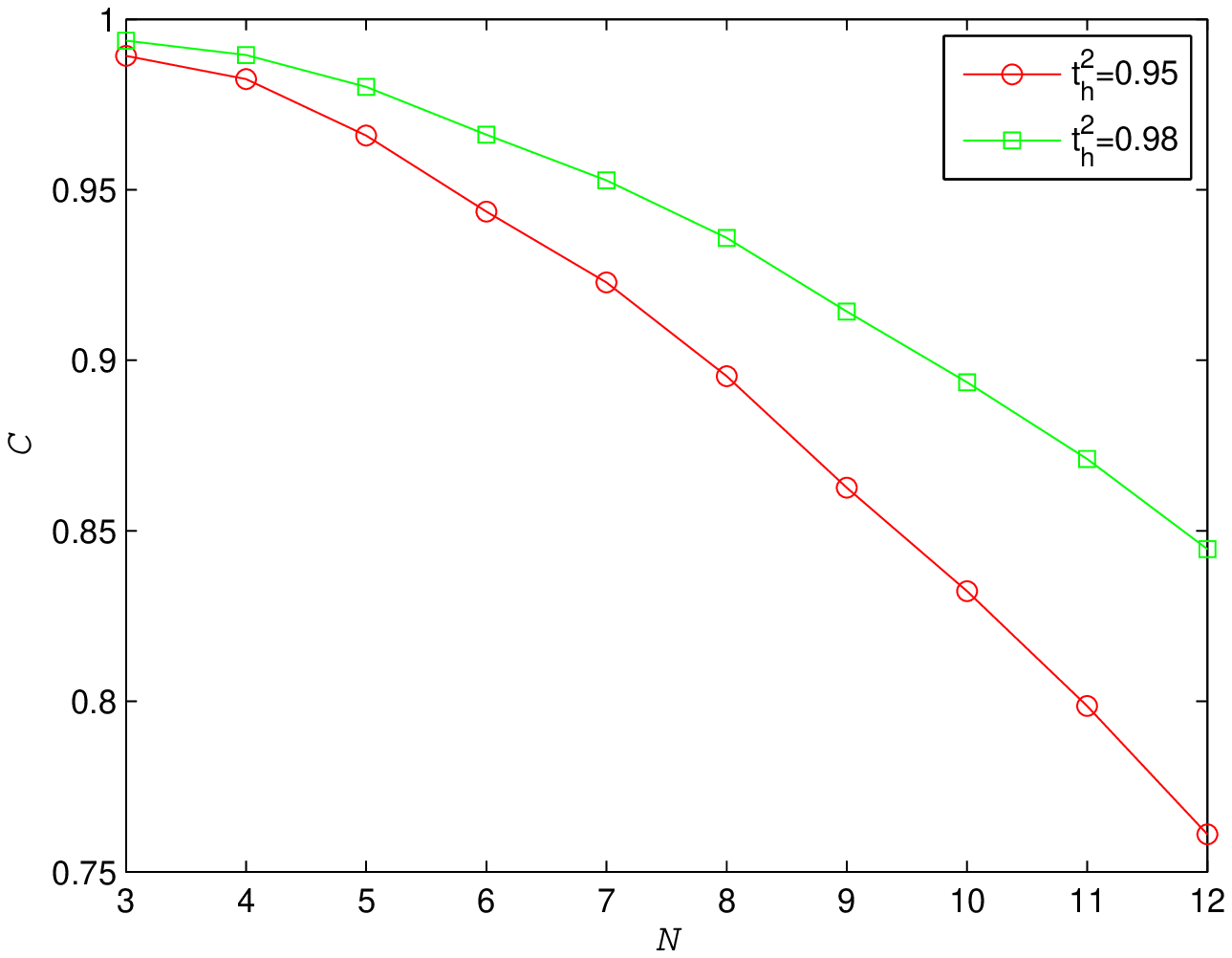}\protect\caption{Imperfect PBS - $\mathcal{\mathcal{C}}_{1,N}$ as a function of the
number of qubits. }
%
%
\label{Fig4}
\end{figure}

\begin{figure}[H]
\centering{}\includegraphics[scale=0.78]{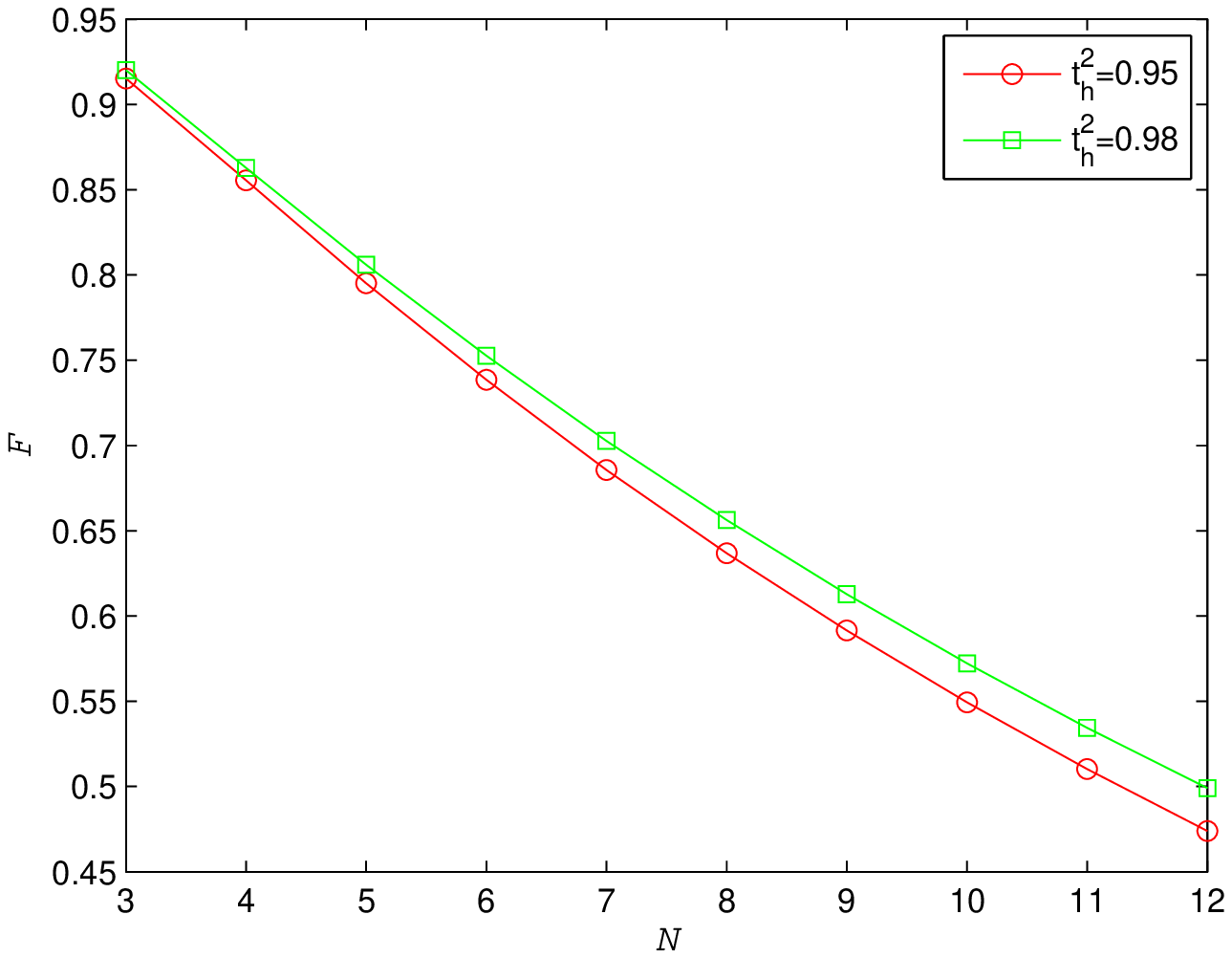}\protect\caption{Imperfect PBS -$\mathcal{F}$ as a function of the number of qubits. }
%
%
\label{Fig5}
\end{figure}

An ideal PBS transmits only horizontally polarized photons and reflects
only vertically polarized photons. In practice, however, horizontally
(vertically) polarized photons are also reflected (transmitted). For
example, a horizontally polarized photon which enters the PBS from
the first input undergoes the transformation
\[
\left|h\right\rangle _{1}\rightarrow t_{h}\left|h\right\rangle _{3}+ir_{h}\left|h\right\rangle _{4},
\]
where the indices $1$ and $2$ ($3$ and $4)$ represent the input
(output) ports. Similarly, a vertically polarized photon which enters
the second input port undergoes the transformation

\[
\left|v\right\rangle _{2}\rightarrow t_{v}\left|v\right\rangle _{4}+ir_{v}\left|v\right\rangle _{3},
\]
 where for an ideal PBS $t_{h}=r_{v}=1$, and $r_{h}=t_{v}=0$. In
our setup we consider the scenario in which one photon enters the
PBS from the first input port and another photon enters the PBS from
the second input port. In addition, the post-selection ensures that
one photon leaves the PBS from the third port and one photon leaves
the PBS from the fourth port. The post-selected operation of the PBS
is therefore given by the two-qubit transformation
\[
T_{PBS}=\left(\begin{array}{cccc}
t_{h}^{2}-r_{h}^{2} & 0 & 0 & 0\\
0 & t_{h}t_{v} & -r_{h}r_{v} & 0\\
0 & -r_{h}r_{v} & t_{h}t_{v} & 0\\
0 & 0 & 0 & t_{v}^{2}-r_{v}^{2}
\end{array}\right),
\]
where the first, second, third and fourth rows correspond to the input
states $\left|h_{1}h_{2}\right\rangle $, $\left|h_{1}v_{2}\right\rangle $,$\left|v_{1}h_{2}\right\rangle $,
and $\left|v_{1}v_{2}\right\rangle $ respectively. For typical PBS
$t_{h}^{2}=0.95$ and $r_{v}^{2}=0.99$, and for high-performance
PBS $t_{h}^{2}=0.98$ and $r_{v}^{2}=0.995$. Plugging these values
in $T_{PBS}$ we obtain $\mathcal{C}_{1,N}$ and $\mathcal{F}$ as
function of the number of qubits (See Fig. \ref{Fig4} and Fig. \ref{Fig5}). The entanglement
lengths of the typical PBS and high-performance PBS are (due to the unitary operation involved -
a negligible amount of entanglement resides, in this
calculation it is addressed with a numerical
threshold of $\mathcal{C}_{1,N}>10^{-2}$, which is set according to measurement accuracy) $\mathcal{L}=75$ and
$\mathcal{L}=185$ respectively. We can therefore conclude that the
imperfection of the PBS is not a limiting factor in our scheme.

\subsection{A two-photon error}

In this subsection we consider the case when the single-photon source
emits two photons and hence, two photons enter the setup and arrive
at the PBS simultaneously. We assume that when two photons arrive
at a detector simultaneously one of the photons is lost and the
other photon is detected (with an equal probability of each photon
to be lost or detected). It is easily verified that when only
one photon enters the loop and two photons arrive at the detector,
the probability of a successful process is $\mathrm{P}_{sp}=2/3$
.For simplicity, we deduce a lower bound on the total success probability
by assuming that whenever two photons enter the loop (except for the
last round %
\footnote{Two photons which enter the loop at the last round also arrive to
a detector so in this case the probability of a successful process
is also given by $\mathrm{P}_{sp}=2/3$. %
}) the probability of a successful process is zero.

\begin{figure}[H]
\centering{}\includegraphics[scale=0.78]{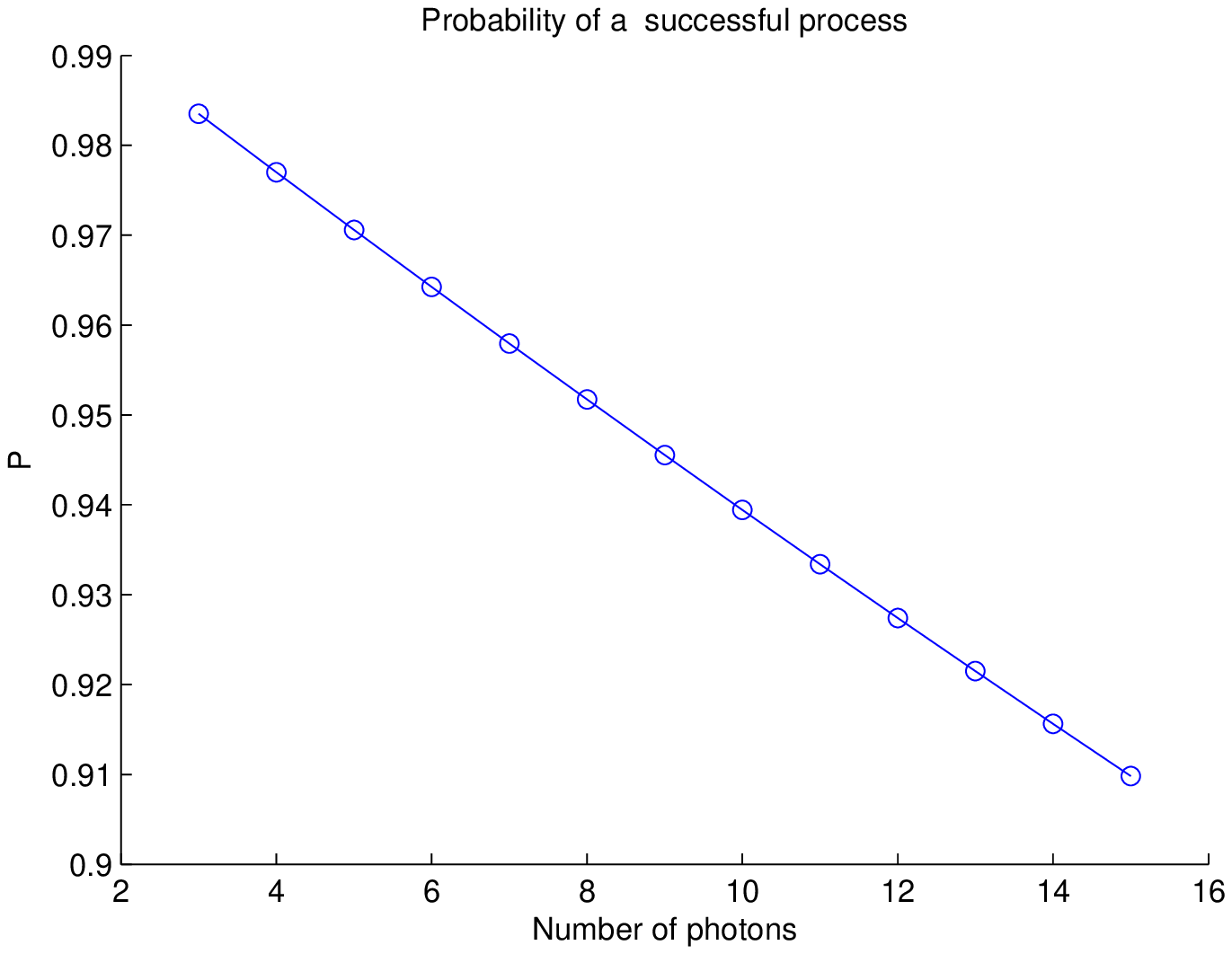}\protect\caption{Lower bound on the probability for a successful generation of an $N$-qubit cluster state
given a non-zero probability ($p=0.01$) for a two-photon emission
by the single-photon source. }
\label{Fig10}
\end{figure}

Denote by $p$ the probability that the single-photon source emits
two photons. For an $N$-photon cluster state, the probability that
two photons will not be emitted at any round is $\left(1-p\right)^{N}$,
and the probability for a two-photon emission in each of the rounds
is therefore $\left(1-\left(1-p\right)^{N}\right)/N$ (assuming the
occurrence of only one event of two-photon emission during the generation
of one cluster-state). Hence, we conclude that the total probability
for a successful process is bounded by
\[
\mathrm{P}_{sp}^{total}\geq\left(1-p\right)^{N}+\frac{\left(1-\left(1-p\right)^{N}\right)\left(N-1\right)}{N}\frac{1}{3}+\frac{\left(1-\left(1-p\right)^{N}\right)}{N}\frac{2}{3}.
\]
 For commonly used single-photon sources $p\approx0.01.$ In Fig.
\ref{Fig10} we plot $\mathrm{P}_{sp}^{total}$ as function of the number of
qubits.

\subsection{CNOT error}

\begin{figure}[H]
\centering{}\includegraphics[scale=0.78]{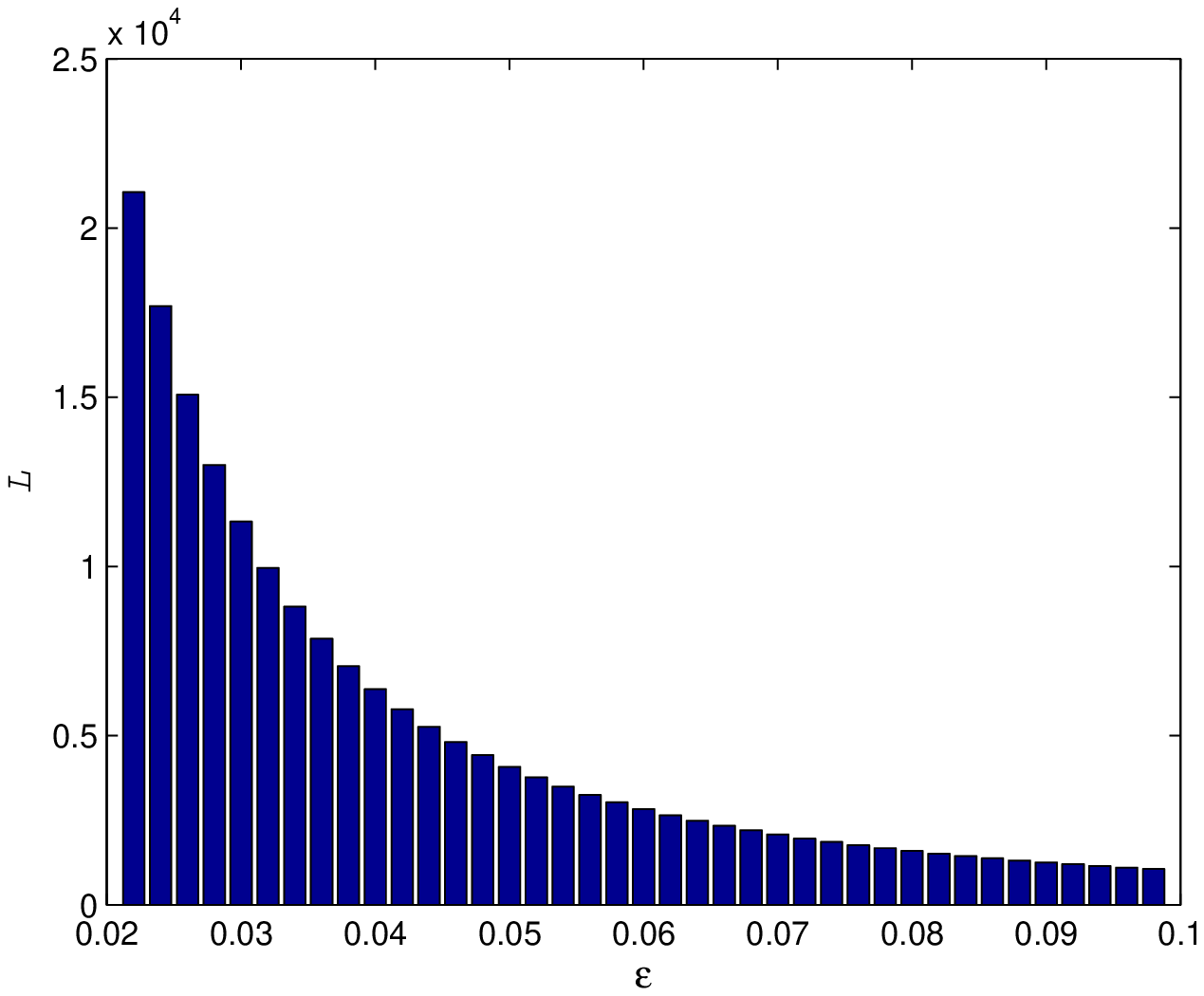}\protect\caption{$\mathcal{L}$ as a function of $\epsilon$ , $0.022\leq\epsilon\leq0.098$
. }
\label{Fig11}
\end{figure}

\begin{figure}[H]
\centering{}\includegraphics[scale=0.78]{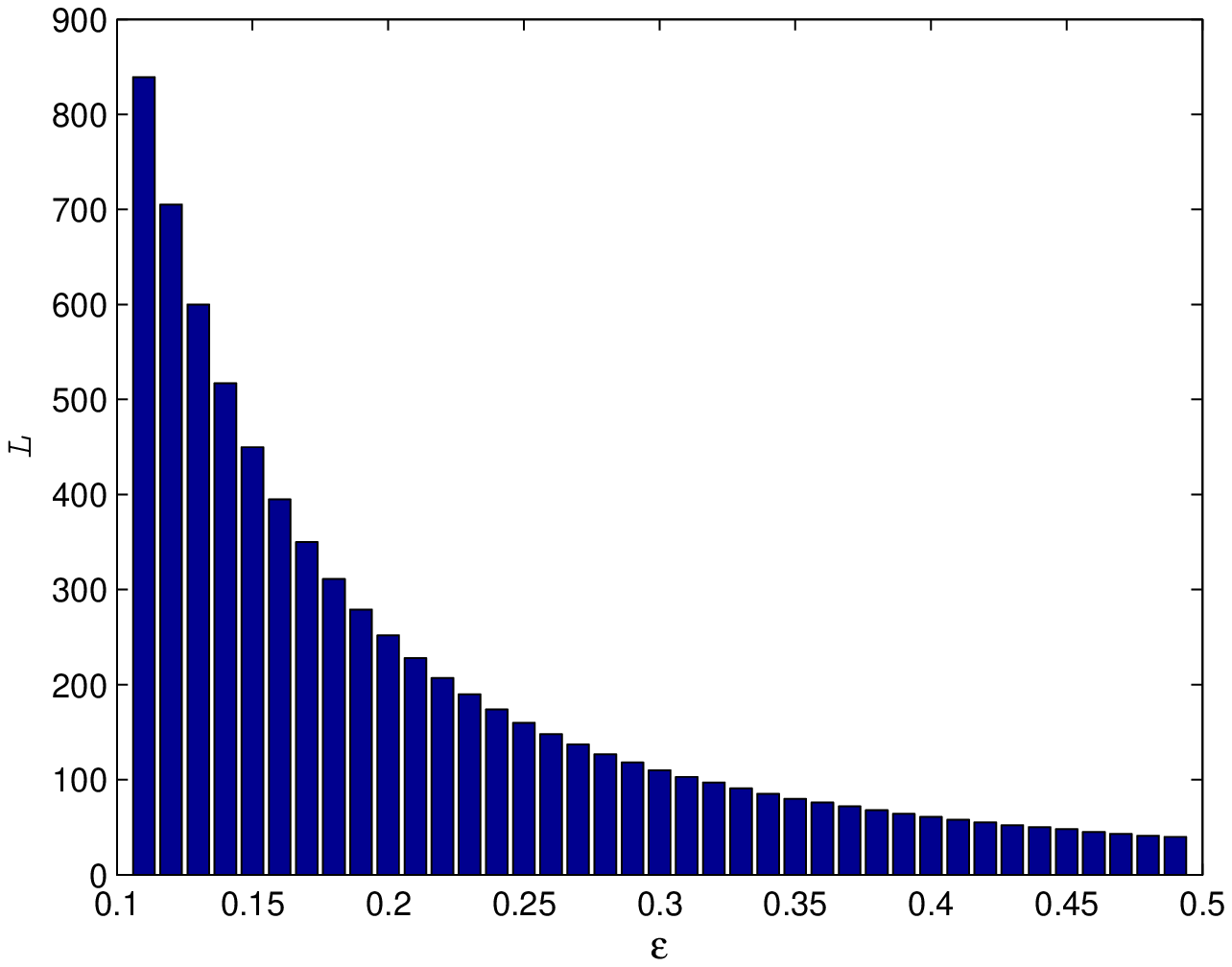}\protect\caption{$\mathcal{L}$ as a function of $\epsilon$ , $0.11\leq\epsilon\leq0.49$
. }
\label{Fig12}
\end{figure}

Most generally, our scheme is formulated such that the photons are
entangled by a CNOT gate (See Fig. 1 in main text). In this section
we consider the effect of an error in the CNOT operation. We model
the CNOT operation by
\[
U_{CNOT}=\begin{pmatrix}1 & 0 & 0 & 0\\
0 & 1 & 0 & 0\\
0 & 0 & \cos\left(\frac{\pi}{2}+\epsilon\right) & i\sin\left(\frac{\pi}{2}+\epsilon\right)\\
0 & 0 & i\sin\left(\frac{\pi}{2}+\epsilon\right) & \cos\left(\frac{\pi}{2}+\epsilon\right)
\end{pmatrix},
\]
where $\epsilon$ represents the error. An ideal CNOT gate is obtained
when $\epsilon=0$. Averaging over $\epsilon$ is obtained by assuming that in each
execution of a CNOT gate the density matrix evolves such that with a probability
of $1/2$ an ideal CNOT is executed, and with a probability of $1/2$ a faulty CNOT with
an error of epsilon is executed (for a given value of $\epsilon$ this corresponds to
an averaging of the error from zero to epsilon). In Fig. \ref{Fig11} and
Fig. \ref{Fig12} we plot $\mathcal{L}$
as a function of $\epsilon$. 

\section{Gate operation}
Following the main text and Methods section, the operation performed is described in the following way.
We first define the Hadamard transform matrix
\[
H=\frac{1}{\sqrt{2}}\left(\begin{array}{cc}
1 & 1\\
1 & -1
\end{array}\right),
\]
and the CNOT operation is
\[
U_{CNOT}=\left(\begin{array}{cccc}
1 & 0 & 0 & 0\\
0 & 1 & 0 & 0\\
0 & 0 & 0 & 1\\
0 & 0 & 1 & 0
\end{array}\right),
\]
The gate operation described in Fig. 1 in the main text is
\[
U_{gate}= \left(H\otimes H\right)^{-1}  U_{CNOT} \left(H\otimes H\right)
\]
so when the first two photons arrive at paths 1 and 3, for example at the state $\left|h_1h_3\right\rangle$
(the other possibilities could be found in the main text),
the resulting state is
\[
\left|\phi\right\rangle=U_{gate} \left( I_2 \otimes H \right) \left|h_1h_3\right\rangle
\]
where
\[
I_2=\left(\begin{array}{cc}
1 & 0\\
0 & 1
\end{array}\right)
\]
This process could be repeated $n$ number of times and produces an entangled linear cluster state of $n+2$ photons.
The resulted $n+2$ $(n>0)$ state is
\[
\left|\phi\right\rangle_{n+2}= \left( U_{gate} \otimes U_I^{\{n\}} \right) \left( \left( I_2 \otimes H \right) \otimes U_I^{\{n\}} \right) \left( \left|h\right\rangle \otimes \left|\phi\right\rangle_{n+1} \right)
\]
where $U_I$ is defined in the following manner
\[
U_I^{\{k+1\}}=I_2 \otimes U_I^{\{k\}}
\]
$U_I^{\{k\}}=1$ for $k=0$.

We present, for clarity, the obtainable states from the gate operation up to $n=3$ (five photon linear cluster state):
\begin{equation}
\begin{aligned}
n=0\Rightarrow \left|\phi\right\rangle_{2}=\ \;\frac{1}{\sqrt{2}}\left( \right.&\left.\left|hh\right\rangle+\left|vv\right\rangle \right)=\left|\phi^+\right\rangle
\\
n=1\Rightarrow \left|\phi\right\rangle_{3}=\ \;\frac{1}{\sqrt{2}}\left( \right.&\left. \left|hhp\right\rangle+\left|vvm\right\rangle\right)=\left|GHZ_3\right\rangle
\\
n=2\Rightarrow \left|\phi\right\rangle_{4}=\ \;\frac{1}{\sqrt{4}}\left( \right.&\left|hhhp\right\rangle+\left|hhvm\right\rangle+\\
                                                                           &\left. \left|vvhp\right\rangle-\left|vvvm\right\rangle\right)
\\
n=3\Rightarrow  \left|\phi\right\rangle_{5}=\ \;\frac{1}{\sqrt{8}}\left( \right.&\left|hhhhp\right\rangle+\left|hhhvm\right\rangle+\left|hhvhp\right\rangle-\left|hhvvm\right\rangle+\\
                                                                             &\left. \left|vvhhp\right\rangle+\left|vvhvm\right\rangle-\left|vvvhp\right\rangle+\left|vvvvm\right\rangle\right)
\end{aligned}
\end{equation}

\section{Experimental interference measurement}
In order to observe the quantum correlation of the
entangled state created, an interference measurement
was performed.
This was obtained by measuring the visibility in
the $\left\{ p,m\right\}$ basis.

For the case of two photon entanglement,where
\[
\left|\phi^+\right\rangle=\frac{1}{\sqrt{2}}\left( \left|hh\right\rangle+\left|vv\right\rangle \right)
\]
is created (the appropriate subscripts denoting each photon
could be found in the main text),
a transformation to the $\left\{ p,m\right\}$ basis would lead to
\[
\left|\phi^+\right\rangle=\frac{1}{\sqrt{2}}\left( \left|pp\right\rangle+\left|mm\right\rangle \right)
\]
Detection in our scheme (see Fig. 3 in main text) comes after a projection
measurement on the $\left\{ h,v\right\}$ basis and four measurement
coincidence outcomes are possible
\[
P_{pp},P_{pm},P_{mp},P_{mm}
\]
which are normalized probabilities (the density matrix diagonal terms $\rho_{ii}$).
In the case of $\left|\phi^+\right\rangle$ the outcome would read
\[
P_{pp}=0.5,P_{pm}=0,P_{mp}=0,P_{mm}=0.5
\]
The 2-photon visibility is defined as
\[
V_2=P_{pp}+P_{mm}-(P_{pm}+P_{mp})
\]

For the case of three photon entanglement, a similar definition of visibility
was required to observe the interference of the state.
Yet, when examining the resulted state
\[
\frac{1}{\sqrt{2}}(|\phi^+h\rangle+|\phi^-v\rangle)=\frac{1}{\sqrt{2}}(|hhp\rangle+|vvm\rangle)
\]
a similar process of transformation, projection and detection would give 8 probabilities
\[
P_{ppp},P_{ppm},P_{pmp},P_{pmm},P_{mpp},P_{mpm},P_{mmp},P_{mmm}
\]
that are all equal, thus the visibility $V_3=0$ and would not be able to
exhibit the existence of quantum correlations as in the two photon case
(this does not occur in linear clusters with higher even number of photons).

Overcoming this problem, without changing the measurement process, is possible
by introducing a birefringent phase $\varphi$ in the delay loop.
The created two photon state is
\[
\left|\phi^{i+}\right\rangle=\frac{1}{\sqrt{2}}\left( \left|hh\right\rangle+e^{i\varphi}\left|vv\right\rangle \right)
\]
following the gate operation on the third photon and an addition of the birefringent phase $\varphi$ in the delay loop, the three photon state is
\[
\frac{1}{\sqrt{2}}(|\phi^{i+} h\rangle+e^{i\varphi}|\phi^{i-}v\rangle)=\frac{1}{\sqrt{4}}(|hhh\rangle+e^{i\varphi}|vvh\rangle+e^{i\varphi}|hhv\rangle-e^{i2\varphi}|vvv\rangle)
\]

Examining the measurement outcome probabilities in the $\left\{ p,m\right\}$ basis gives two different dependencies on $\varphi$.
For $P_{ppp},P_{pmm},P_{mpm},P_{mmp}$ (odd number of $p$) the dependence is
\[
\frac{1}{32}(6-2cos(2\varphi))
\]
while for $P_{ppm},P_{pmp},P_{mpp},P_{mmm}$ (even number of $p$) it is
\[
\frac{1}{32}(2+2cos(2\varphi))
\]
It is clear that for the case of $\varphi=0^{\circ}$ the two terms are equal, while for
$\varphi=90^{\circ}$ the first (second) term interferes constructively (destructively).
In this way the three photon visibility is defined as
\[
V_3=P_{ppp}+P_{pmm}+P_{mpm}+P_{mmp}-(P_{ppm}+P_{pmp}+P_{mpp}+P_{mmm})
\]

Notice, that for the case of $\varphi=90^{\circ}$, the 2-photon visibility $V_2=0$.